\documentclass[11pt]{article}
\usepackage[left=1in,top=1in,right=1in,bottom=1in]{geometry}
\usepackage{times}

\usepackage[round]{natbib}

\usepackage{url}
\usepackage{verbatim}
\usepackage{amsmath}
\usepackage{amssymb}
\usepackage{amsthm}
\usepackage{rotating}
\usepackage{algorithm}
\usepackage[noend]{algpseudocode}

\usepackage{tabularx}
\usepackage{subfigure}
\usepackage{multirow}
\usepackage{siunitx}

\begin{document}
\title{Prediction of Bayesian Intervals for Tropical Storms}

\author{Max Chiswick, Sam Ganzfried$^{1}$\\ 
$^1$Ganzfried Research
}

\date{\vspace{-5ex}}

\maketitle
\begin{abstract}
Building on recent research for prediction of hurricane trajectories using recurrent neural networks (RNNs), we have developed improved methods and generalized the approach to predict Bayesian intervals in addition to simple point estimates. Tropical storms are capable of causing severe damage, so accurately predicting their trajectories can bring significant benefits to cities and lives, especially as they grow more intense due to climate change effects. By implementing the Bayesian interval using dropout in an RNN, we improve the actionability of the predictions, for example by estimating the areas to evacuate in the landfall region. We used an RNN to predict the trajectory of the storms at 6-hour intervals. We used latitude, longitude, windspeed, and pressure features from a Statistical Hurricane Intensity Prediction Scheme (SHIPS) dataset of about 500 tropical storms in the Atlantic Ocean. Our results show how neural network dropout values affect predictions and intervals. 
\end{abstract}

\section{Introduction}
We look at a dataset of tropical storm data in the Atlantic Ocean from 1982 to 2017 and perform deep learning predictions with uncertainty bounds on trajectories of the storms. The result of these storms, particularly the strongest ones called \emph{hurricanes}---defined as having wind speeds exceeding 74 mph---can be devastating because of their strong winds and heavy precipitation that can cause dangerous tides. Tropical storms can cause major environmental disasters when they reach land, such as the 2005 Hurricane Katrina that resulted in over 850 deaths and caused major economic damage and the 2012 Hurricane Sandy that caused almost \$70 billion in damage across much of the eastern United States, with peak winds of 115 mph~\citep{Sandy13}. According to the National Oceanic and Atmospheric Administration, it is likely that global warming will cause hurricanes in the upcoming century to be more intense by 1 to 10\% globally (with higher peak winds and lower central pressures), which will result in a higher proportion of more severe storms~\citep{NOAA19:Global}. 

Historically, hurricane trajectory predictions have used statistical methods that can be limiting because of the nonlinearity and complexity of atmospheric systems. Deep learning techniques and specifically recurrent neural networks have grown in popularity in recent years as a strong method for approaching prediction problems because of the ability to extract important features and relationships from complex high-dimensional data, especially for forecasting and classification~\citep{McDermott17:Bayesian}. We implemented a number of improvements over previous deep learning prediction work~\citep{Alemany19:Predicting}, including predicting exact storm locations in latitude/longitude instead of a grid value and using a prediction window that uses all previous hurricane data rather than a fixed-size sliding window. 

While hurricane trajectory predictions have seen improvements recently~\citep{SHIPS}, we build on previous work to include a fundamental uncertainty measure in the prediction for the first time as part of a neural network framework. The uncertainty measure is especially valuable for understanding a defined location range rather than only a point estimate, which is important for evacuation and safety/preparation purposes. The National Hurricane Center (NHC) builds their uncertainty cone such that $\frac{2}{3}$ of historical forecast errors over the previous 5 years fall within the circle, whereas we use a rigorous Bayesian prediction model to build our intervals. 

\section{RNNs and Dropout}
RNNs are fully connected networks, which use connection weights as training parameters. The standard RNN setup uses inputs over time that are connected to hidden layers. The hidden layers connect forward to the next hidden layer or output layer and also through time to the next hidden layer timestep. The $n_y$ dimensional output vector $Y_t$ corresponds to the original $n_x$ dimensional input vector $X_t$ by:
$$Y_t = g(V*h_t)$$
where $h_t$ is the final $n_h$ dimensional vector of hidden state variables, $V$ is the $n_y \times n_h$ weight matrix, and the function $g(\cdot)$ is an activation function that maps between the output and hidden states. The hidden layers are defined as follows:
$$h_t = f(W*h_{t-1} + U*X_t) $$
where $W$ is an $n_h \times n_h$ weight matrix, $U$ is an $n_h \times n_x$ weight matrix, and the function $f(\cdot)$ is the activation function for the hidden layers, which creates the nonlinearity in the neural network. 

Deep learning methods are known to overfit, which results in an inability of the model to generalize properly. We use \textit{dropout} to regularize the network and prevent this overfitting. Dropping neurons randomly during training is known to reduce the generalization error.

By making the presence of hidden units unreliable, dropout prevents co-adaptions amongst the nodes, and promotes each to be more robust and to produce more useful features on its own without relying on other hidden units~\citep{Srivastava14:Dropout}. Dropout is a common technique in which a hyperparameter with a set percentage is given as the percentage number of neurons to set to 0 (i.e., to dropout) during the training passes. Large dropout rates can lead to divergence, while small rates can result in insignificant effects. In the next section we show that additionally using dropout during training approximates Bayesian inference in a deep Gaussian process. 

\citet{Srivastava14:Dropout} note that in the simplest case, each hidden unit is retained with a fixed probability $p$ independent of other units, where $p$ can be chosen using a validation set or can simply be set at 0.5, which seems to be close to optimal for a wide range of networks and tasks. For the input units, however, the optimal probability of retention is ``usually closer to 1 than to 0.5.''

We used Long Short-Term Memory Cells (LSTM) in our RNN, whose main purpose is to remember values over arbitrary time intervals by preventing vanishing and exploding weights throughout the RNN. LSTMs have been shown to significantly improve RNN performance when applications require long-range reasoning. Storm models that span up to 120 hours in length are a good fit for the LSTM model~\citep{Alemany19:Predicting}. 

In general, estimating posterior densities of weights using Bayes' rule is difficult because of the need to marginalize over all possible values that the weight parameter can take in the model. A Gaussian prior on the weights is generally used, $p(w) = N(0,1)$. 

$$ p(w \mid x,y) = \frac{p(x,y \mid w) \, p(w)}{\int p(x,y \mid w) \, p(w) dw} $$

This process is equivalent to variational inference in Gaussian processes; i.e., by averaging the forward passes through the network, this is equivalent to Monte Carlo integration over a Gaussian process posterior approximation. Variational inference is an approach for approximating a model's posterior that would be otherwise difficult to work with directly. By minimizing the Kulback-Leibler (KL) divergence between an approximating variational distribution $q_\theta(w)$ and $p(w \mid x,y)$, we can estimate our original predictive distribution for new input $x^*$ 
$$p(y^* \mid x^*,X,Y) = \int p(y^* \mid x^*,w) p(w \mid X,Y) dw$$ 
$$\mbox{to }q_\theta(y^* \mid x^*) = \int p(y^* \mid x^*, w) q_\theta(w) dw,$$ 
which can be approximated at prediction time by 
$$q_\theta(y^* \mid x^*) \approx \frac{1}{T} \sum_{t=1}^{T} p(y^* \mid x^*,w_t).$$

\section{Quantifying uncertainty with a Bayesian RNN}
Prior research has shown that we can use a deep learning model that uses dropout to model uncertainty by using the dropout in both the training and testing passes, without the need for an inference framework~\citep{Gal16:Dropout,Gal16:Theoretically,Gal16:Uncertainty}.  This technique of interpreting dropout as a Bayesian approximation of a deep Gaussian process provides a simple method for estimating the interval of a neural network output, in addition to the standard point estimate output. The key idea with Bayesian RNN dropout is that weights have probability distributions rather than fixed point estimates. We perform multiple forward passes of the network, each time with a new set of weights which result from a different set of dropouts during the prediction phase. 

\begin{figure*}[!ht]
\centering

\begin{minipage}{.51\textwidth}
\centering
\includegraphics[width=\linewidth]{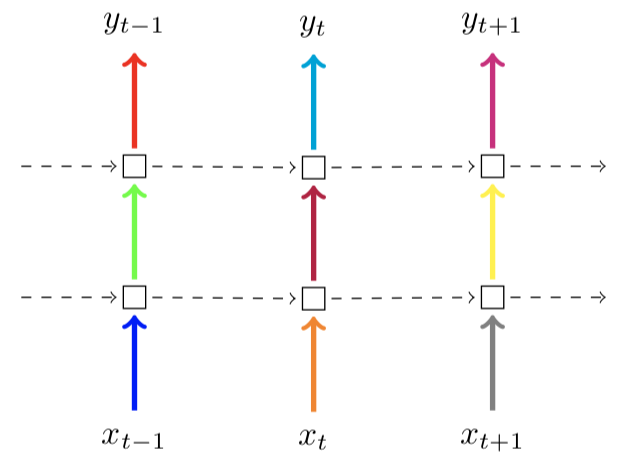}
 \caption{Standard RNN model}
 \label{fi:naivernn}
\end{minipage}
\hfill
\begin{minipage}{.48\textwidth}
\centering
\includegraphics[width=\linewidth]{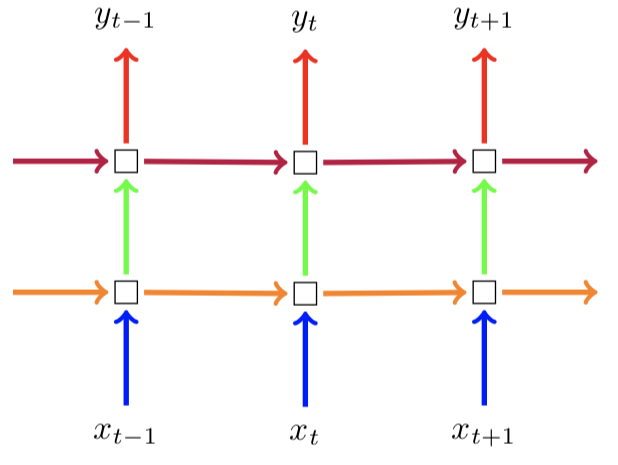}
 \caption{Variational Bayes RNN model}
 \label{fi:variationalrnn}
\end{minipage}
\end{figure*}

The authors showed that if dropout is seen as a variational Monte Carlo approximation to a Bayesian posterior, then the natural way to apply it to recurrent layers is to generate a dropout mask that zeroes out both feedforward and recurrent connections for each training sequence, but to keep the same mask for each time step in the sequence as a form of variational inference (see Figures~\ref{fi:naivernn} and~\ref{fi:variationalrnn}). This differs from the na\"ive way of applying dropout to RNNs, which would generate new dropout masks for each input sample regardless of which time sequence it was from. 

By enabling dropout during the testing phase, every forward pass with a given input will result in a different output. These non-deterministic predictions can be interpreted as samples from a probabilistic distribution, i.e., a Bayesian approximation. By applying dropout to all the weight layers in a neural network, we are essentially drawing each weight from a Bernoulli distribution. This means that we can sample from the distribution by running several forward passes through the network. By sampling from distributions of weights, we can evaluate the distribution of many predictions, which informs the quality of the model and the uncertainty of the output. The more the output varies when using dropout, the higher the model's output uncertainty. The sample means and variances of the output represent estimates of the predictive mean and variance of the model. 

\begin{figure}[!ht]
\centering
  \includegraphics[width=0.9\linewidth]{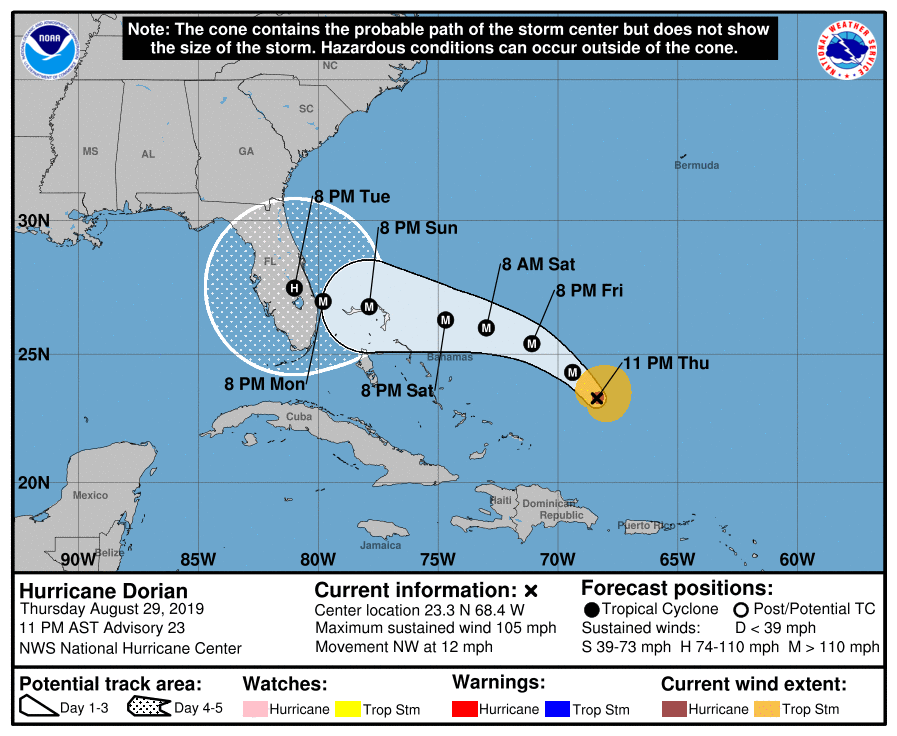}
 \caption{NHC Forecast Cone for Hurricane Dorian (2019)}
 \label{fi:cone}
\end{figure}

The NHC publishes a ``Forecast Cone'' that shows estimates for two times (e.g., 8AM and 8PM) for each future day up to 5 days ahead (Figure~\ref{fi:cone}). It shows a 67\% interval around the center of the storm for each of those times. It is named a cone because the areas closest to the present time are relatively thin due to having less uncertainty and the times further in the future are wider due to having more uncertainty. The two main difficulties with this diagram style are that hurricanes can be hundreds of miles wide, potentially wider than the cone itself, and that by definition of the 67\% uncertainty interval, about $\frac{1}{3}$ of hurricanes will be outside of the cone~\citep{Cairo19:Those}. 

The size of the circles is set so 67\% of historical official forecast errors over the previous 5-year sample fall within the circle. For 2019, a 24-hour forecast of an Atlantic hurricane would have a 41 nautical mile probability circle. Our uncertainty interval instead uses fundamental Bayesian techniques by performing multiple predictions using our model and evaluating according to the mean and variance of these predictions. Additionally, we examine a variety of interval ranges up to 99\%, which could be more useful since fewer hurricanes would be expected to be outside of the interval. 

\section{Applying the model to Atlantic hurricanes}
We used the SHIPS dataset with data from Atlantic Ocean storms from 1982 to 2017~\citep{SHIPS}. The data contains 499 storms, with a large number of time series records for each storm and about 100 total features (not all are available for all storms). Each data point is given in six-hour timestep intervals. The base features that we use are latitude, longitude, maximum surface wind (kt), minimum sea level pressure (hPa), and calculated features of distance and direction (angle of travel) from timestep to timestep. All fields are discretized relative to the storm center, which was determined from the NHC best track. All atmospheric predictors are from the NCEP global model analyses. 

Our model takes an input in the shape of the number of samples by the maximum hurricane length by the number of features (6). We use 2 LSTM hidden layers with sizes of 32 and then 16, which isolate important hidden dynamic features from the input data. Our output for the trajectory predictions are latitude and longitude at the appropriate prediction timestep. Loss is computed using mean squared error (MSE), which takes our predicted value for the location ($\hat{y}$) and the labeled known value ($y$) and computes $\text{MSE} = (y - \hat{y})^2$. This is averaged over all predictions in the training dataset. 

To implement this model, we use Keras and the parameter \textit{dropout} for input dropout W weights and \textit{recurrent\_dropout} for hidden state dropout (U weights). Keras is automatically set up to keep the same dropout masks over all timesteps as required by the Bayesian model. We trained over 200 epochs, at which point the model stopped improving. 

All hurricanes in the dataset were shuffled (i.e., each hurricane remained complete, but the order of hurricanes was randomized) to ensure that the training and test sets don't contain only older or newer data. All features are then min-max normalized prior to training. 

We build our training and test set input with 3 main parameters: \textit{min\_to\_start\_predictions}, \textit{prediction\_length}, and \textit{max\_hurricanes}. The first parameter determines how many preliminary timesteps are seen before any predictions begin. This is used so all predictions have some reasonable basis and is justified because we assume some time is needed to detect the storm. The second is the length of the predictions beyond the input data (e.g., 1 timestep is a 6-hour prediction using all previous input data and 4 timesteps is a 24-hour prediction). The final parameter is fixed at the length of the longest hurricane (89 timesteps in this dataset), such that each model input is this length minus the prediction length minus the minimum prediction start, and shorter inputs are padded with 0s. The labeled data is taken from the corresponding value being predicted (latitude and longitude) at the timestep of the prediction. The inputs and labels are constructed for each hurricane separately and then we input all the data into the network to evaluate the overall error rates. 

We used the following parameters for our experiments:
\begin{itemize}
\item Test set size: 0.25 
\item Validation set size: 0.25
\item Batch size: 64
\item Epochs: 200
\item Optimizer: Adam with default parameters (LR 0.001)
\item Recurrent dropout: 0.1
\item Dropout: Experiments with 0.1/0.2/0.5
\end{itemize}

After model training, we perform the prediction on our test set many times---we used 100 and 400 comparisons for our predictions. Each contains a matrix of predictions over every timestep in the test set and each prediction is unique because of the dropout during testing. We use the D'Agostino-Pearson Test that combines skew and kurtosis to produce an omnibus test of normality and determine that the predictions are approximately normal. We then use the mean and standard deviation of the predictions as our uncertainty measure, allowing us to create uncertainty bounds by using Z-scores for 67\%, 90\%, 95\%, 98\%, and 99\% credible intervals. Credible intervals are a Bayesian concept that denotes that our intervals are based on sampling the network, different from the confidence interval that would indicate knowing a true parameter and creating an interval to include that parameter with some minimum probability. 

\begin{figure*}[!ht]
\centering

\begin{minipage}{.48\textwidth}
\centering
\includegraphics[width=\linewidth]{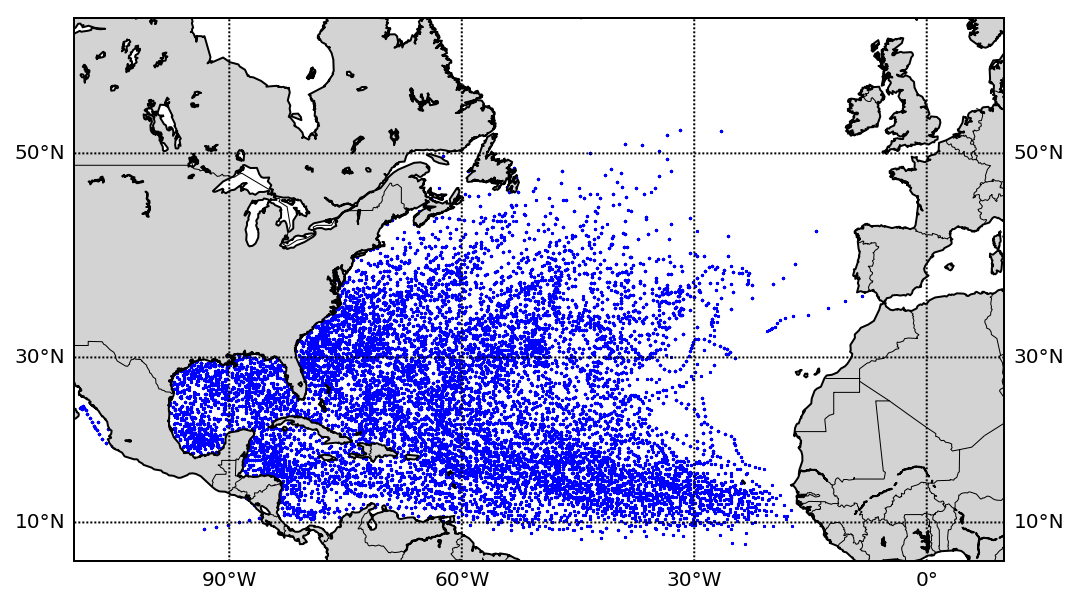}
 \caption{Datapoints from all storms in the dataset}
 \label{fi:alldatapoints}
\end{minipage}
\hfill
\begin{minipage}{.51\textwidth}
\centering
\includegraphics[width=0.94\linewidth]{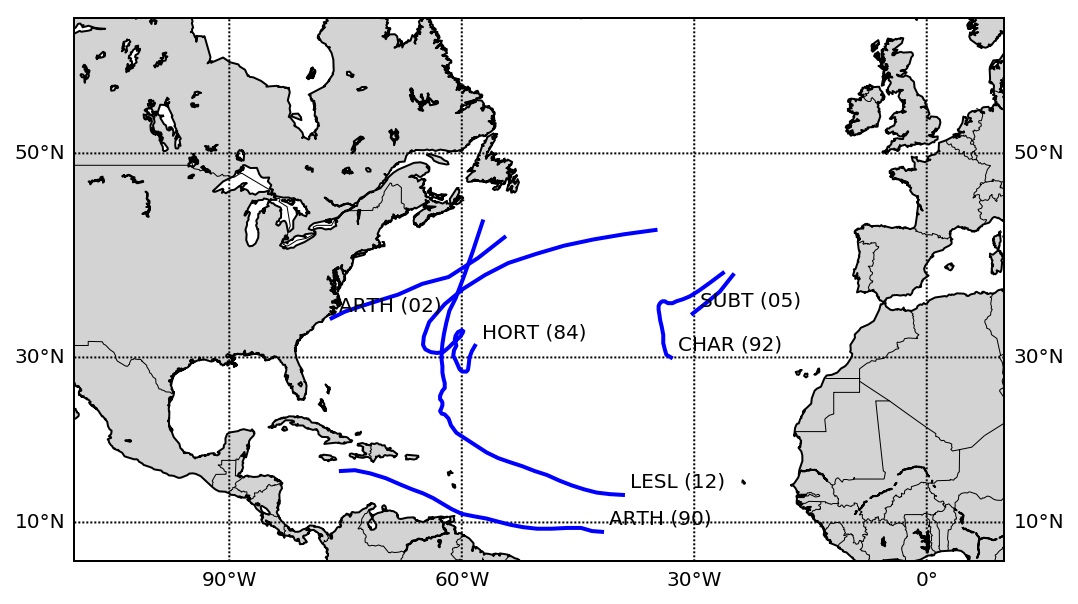}
 \caption{Random selection of 6 hurricanes in dataset}
 \label{fi:randomhurricanes}
\end{minipage}
\end{figure*}

All datapoints from storms in the dataset are shown in Figure~\ref{fi:alldatapoints} and six randomly selected storms are shown for illustration purposes on a map in Figure~\ref{fi:randomhurricanes}.

\section{Experiments}
Our main experiments are predicting locations and intervals of these predictions over the SHIPS dataset. The experiments are over our entire test set and we have also produced figures for Hurricane Katrina specifically. 

\subsection{Trajectory forecast results}
Based on predicting normalized latitude and longitude points, we find the following MSE rates on our test data for three dropout rates in Table~\ref{ta:predictionsmse}. 

\begin{table}[!ht]
\centering
\begin{tabular}{|*{5}{c|}} \hline
\textbf{0.1 dropout} &\textbf{0.2 dropout} &\textbf{0.5 dropout} \\ \hline
.0020 &.0027 &.0055 \\ \hline
\end{tabular}
\caption{MSE of our model at different dropout levels.}
\label{ta:predictionsmse}
\end{table}

\begin{figure}[!ht]
\centering
  \includegraphics[width=0.9\linewidth]{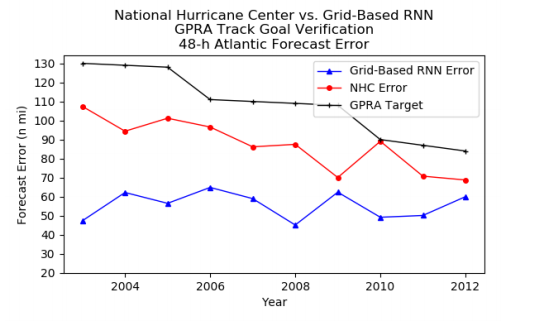}
 \caption{RNN forecast error compared to NHC and GPRA target}
 \label{fi:acomp}
\end{figure}

We performed MSE calculations specifically on Katrina and Sandy. With dropout of 0.2, we had MSE .0017 for Sandy and .0026 for Katrina. We find that higher dropout levels generally resulted in worse MSE, suggesting that the model is not overfitting. \citet{Alemany19:Predicting} used the same RNN prediction techniques and show improvement relative to the National Hurricane Center (NHC) error rates and Government Performance and Results Act (GPRA) target rates for 2003--12 (Figure~\ref{fi:acomp}).

\subsection{Analysis of uncertainty in trajectory forecasts}
\begin{figure*}[!ht]
\centering
 \includegraphics[width=0.9\linewidth]{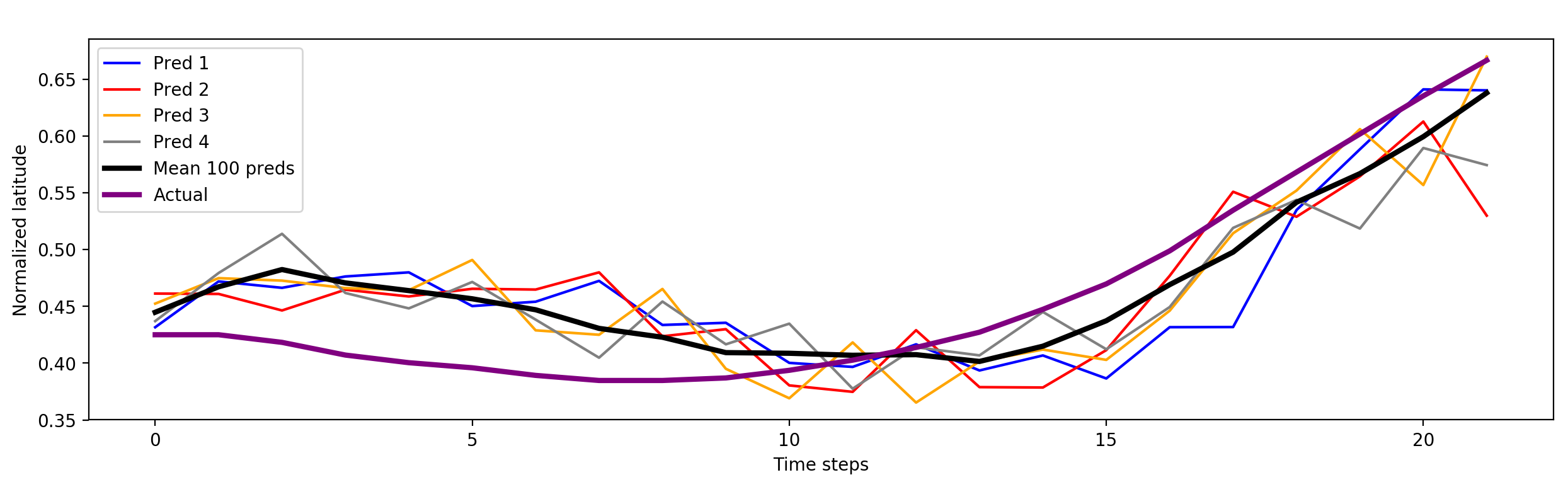}
 \caption{Hurricane Katrina latitude predictions}
 \label{fi:latitude-predictions-kat}

 \includegraphics[width=0.9\linewidth]{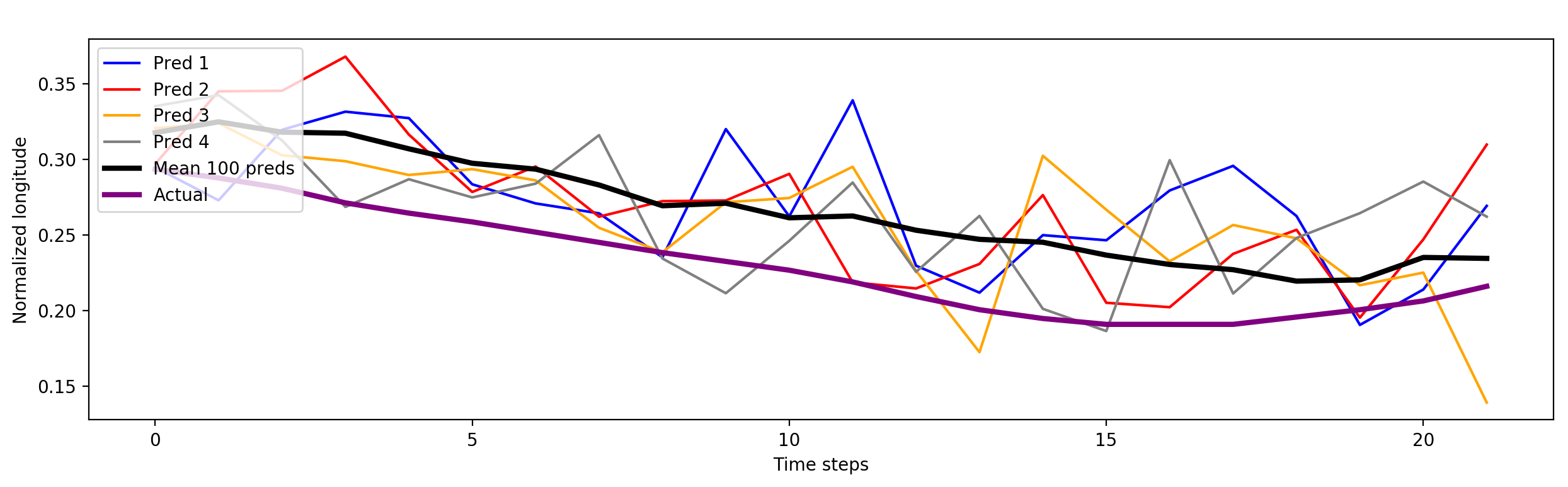}
 \caption{Hurricane Katrina longitude predictions}
 \label{fi:longitude-predictions-kat}

  \includegraphics[width=0.9\linewidth]{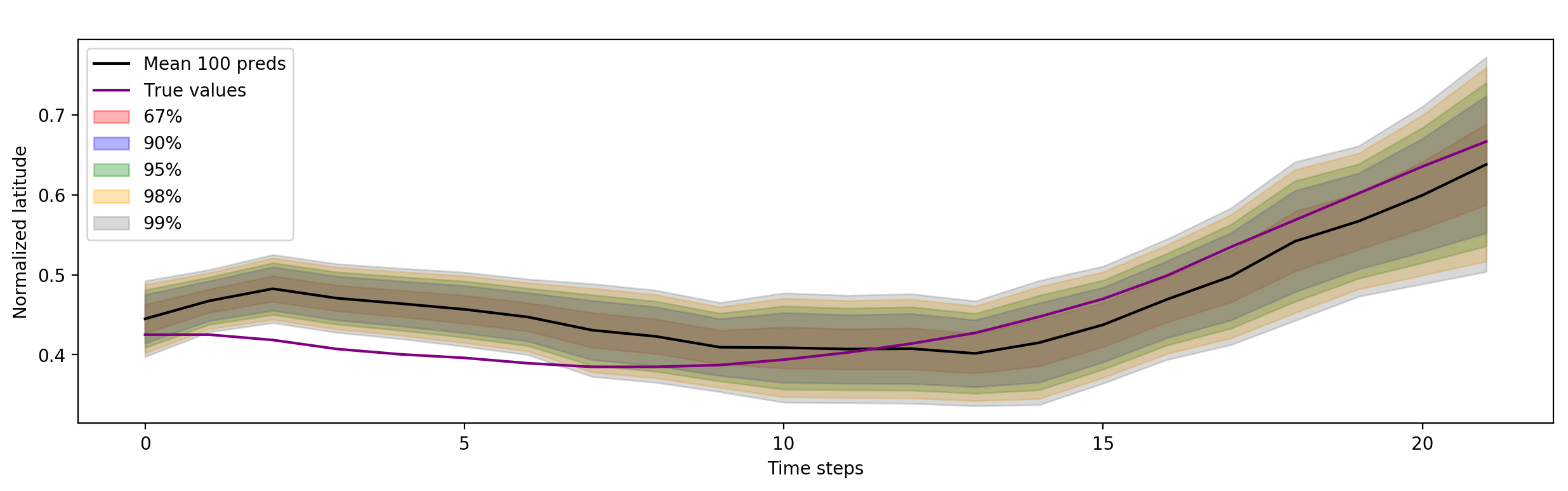}
 \caption{Hurricane Katrina latitude intervals}
 \label{fi:latitude-intervals-kat}

  \includegraphics[width=0.9\linewidth]{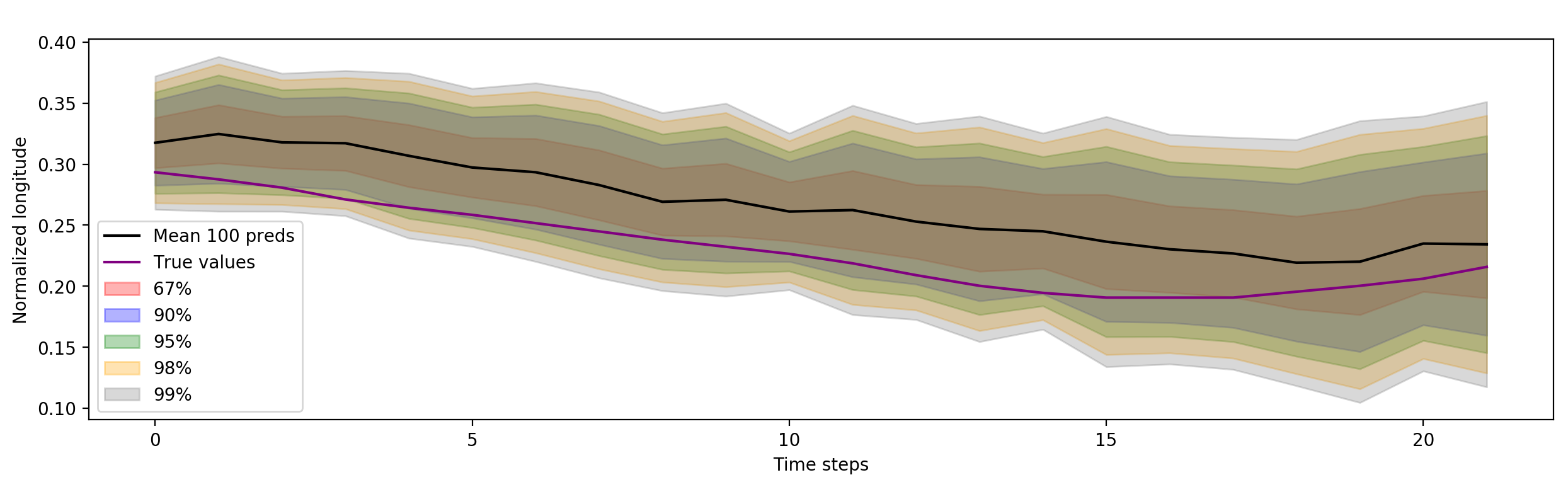}
 \caption{Hurricane Katrina longitude intervals}
 \label{fi:longitude-intervals-kat}
\end{figure*}

We performed uncertainty forecasts on the test portion of our dataset and on two specific hurricanes. Figures~\ref{fi:latitude-predictions-kat}--\ref{fi:longitude-predictions-kat} show 4 random predictions out of 100 total predictions made, along with the mean of the 100 predictions and the true values. This illustrates the variability in the individual predictions. We show one figure each for latitude and longitude for Hurricane Katrina. We see that the dropout does cause noticeable differences in the random predictions. Next we show intervals for latitude and longitude from 100 predictions using their mean, standard deviation, and the relevant Z-score. We show lines for the true values and the mean of our predictions, along with bounds on 67\%, 90\%, 95\%, 98\%, and 99\% intervals on Katrina (Figures~\ref{fi:latitude-intervals-kat}--\ref{fi:longitude-intervals-kat}).

Our evaluation metric for the intervals is to compute the percentage of points over a specific sample or specific hurricane that actually fit within each of the computed interval bands over every timestep of that sample or hurricane. For the main test set we also performed an experiment with 400 predictions in addition to 100, but the effect was negligible. However, increased dropout, which produces more variance in the predictions (although gives the overall model a worse MSE) leads to significantly improved uncertainty estimates. Tables~\ref{ta:predictions1}--~\ref{ta:predictions5} show results using different dropout rates and number of predictions over the entire test set. 

\renewcommand{\tabcolsep}{4pt}
\begin{table}[!ht]
\centering
\begin{tabular}{|*{6}{c|}} \hline
\textbf{0.1 dropout test set} &\textbf{67\%} &\textbf{90\%} &\textbf{95\%} &\textbf{98\%} &\textbf{99\%} \\ \hline
100 Latitude &51.0 & 72.0 &78.5 &84.3 &87.0 \\ \hline
400 Latitude &51.7 & 72.1 &78.6 &84.8 &87.4 \\ \hline
100 Longitude &64.5 & 83.1 &87.3 &90.7 &92.0 \\ \hline
400 Longitude &64.9 & 83.5 &87.5 &90.7 &92.2 \\ \hline
\end{tabular}
\caption{Percentage of latitude and longitude data within predicted region for different interval levels and numbers of predictions using 0.1 dropout.}
\label{ta:predictions1}

\begin{tabular}{|*{6}{c|}} \hline
\textbf{0.2 dropout test set} &\textbf{67\%} &\textbf{90\%} &\textbf{95\%} &\textbf{98\%} &\textbf{99\%}\\ \hline
100 Latitude &61.1 & 82.0 &87.0 &90.9 &93.0 \\ \hline
400 Latitude &61.2 & 82.4 &87.3 &91.2 &93.4 \\ \hline
100 Longitude &66.3 & 84.3 &88.6 &92.2 &93.9 \\ \hline
400 Longitude &66.2 & 84.6 &88.8 &92.0 &94.1 \\ \hline
\end{tabular}
\caption{Percentage of latitude and longitude data within predicted region for different interval levels and numbers of predictions using 0.2 dropout.}
\label{ta:predictions2}

\begin{tabular}{|*{6}{c|}} \hline
\textbf{0.5 dropout test set} &\textbf{67\%} &\textbf{90\%} &\textbf{95\%} &\textbf{98\%} &\textbf{99\%}\\ \hline
100 Latitude &76.4 & 91.7 &94.2 &96.2 &97.0 \\ \hline
400 Latitude &77.1 & 92.0 &94.3 &96.2 &97.1 \\ \hline
100 Longitude &82.0 & 94.3 &95.9 &97.5 &98.0 \\ \hline
400 Longitude &82.5 & 94.2 &96.2 &97.7 &98.2 \\ \hline
\end{tabular}
\caption{Percentage of latitude and longitude data within predicted region for different interval levels and numbers of predictions using 0.5 dropout.}
\label{ta:predictions5}
\end{table}

The 0.1 dropout rate underfits the intervals due to a lack of variation in the predictions. The 0.5 dropout rate produces worse results and has a large variance in the predictions such that the intervals are too large, even larger than the intended sizes (e.g., 100 latitude predictions fits 76.4\% of data rather than 67\%). We determine that 0.2 is the best medium that results in strong error rates and accurate intervals and use this dropout for our individual hurricane predictions. 

\section{Conclusion}
We produced a storm prediction model capable of strong trajectory predictions. By implementing a rigorous uncertainty bound, we add significant value relative to point estimates that were predicted in previous work and an important alternative to the uncertainty bounds produced by the NHC that are based on recent historical data. The computed bounds agree closely with our predicted intervals.

Future work could examine changes in hurricane trajectories over time given the variation in climate change and weather patterns. It may be important to build models based only on more recent hurricanes that best capture the most recent climate effects. We could also utilize more features from the SHIPS dataset. While there are nearly 100 features, many are not provided for large sets of storms. It would be useful to work with a domain expert to understand which features would be most interesting and also if we could further improve our model by incorporating additional weather-related data. If possible to supplement the dataset with satellite imagery, this could be a powerful combination. Another possibility is to compare uncertainty measures using dropout as we have done in this paper with a Bayesian RNN model trained with traditional Markov Chain Monte Carlo methods such as the one described formally by~\citet{McDermott17:Bayesian}. Finally, it could also be useful to perform more specific predictions relating to when and where hurricanes are expected to hit land. 

\section{Acknowledgments}
We would like to gratefully acknowledge the assistance and mentorship from Mustafa K. Mustafa, given as part of the ICLR Climate Change Mentorship Program. We also thank Sheila Alemany for providing code and detailed descriptions of a prior approach for using an RNN to predict hurricane trajectories~\citep{Alemany19:Predicting}, as well as for suggesting the prediction of Bayesian intervals as a future direction.

\clearpage
\bibliographystyle{plainnat}
\bibliography{C://FromBackup/Research/refs/dairefs}

\end{document}